\RequirePackage{fix-cm}
\documentclass[twocolumn]{svjour3}          
\smartqed  
\usepackage{multirow}
\usepackage{subfigure}
\usepackage{graphicx}
\begin{document}

\title{Neuron inspired data encoding memristive multi-level memory cell
}


\author{Aidana Irmanova         \and
        Alex Pappachen James 
}


\institute{F. Author \at
        Nazarbayev University\\
        \email{aidana.irmanova@nu.edu.kz}
             \and
           S. Author \at
          Nazarbayev University\\
                 \email{apj@ieee.com}
}


\maketitle

\begin{abstract}
Mapping neuro-inspired algorithms to sensor backplanes of on-chip hardware require shifting the signal processing from digital to the analog domain, demanding memory technologies beyond conventional CMOS binary storage units. Using memristors for building analog data storage is one of the promising approaches amongst emerging non-volatile memory technologies. Recently, a memristive multi-level memory (MLM) cell for storing discrete analog values has been developed in which memory system is implemented combining memristors in voltage divider configuration. In given example, the memory cell of 3 sub-cells with a memristor in each was programmed to store ternary bits which overall achieved 10 and 27 discrete voltage levels. However, for further use of proposed memory cell in analog signal processing circuits data encoder is required to generate control voltages for programming memristors to store discrete analog values. In this paper, we present the design and performance analysis of data encoder that generates write pattern signals for 10 level memristive memory.  
\end{abstract}

\section{Introduction}
\label{intro}

\begin{figure*}[ht!]
\includegraphics[width=0.8\textwidth]{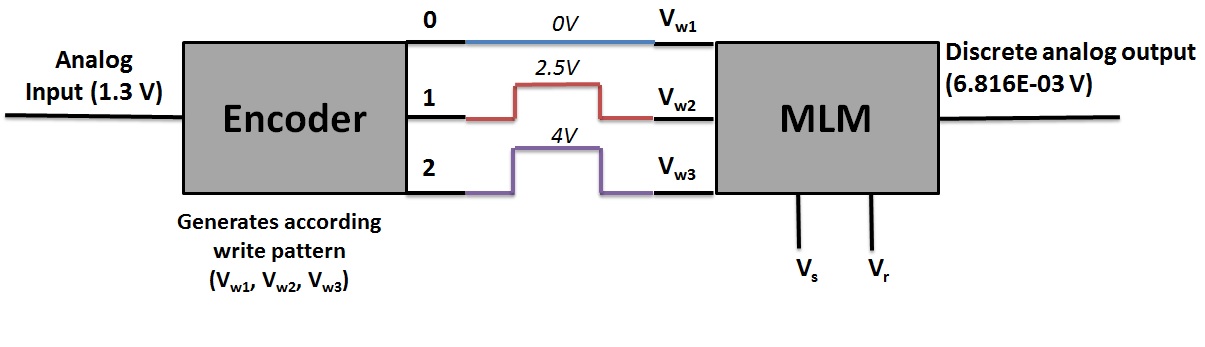}
\centering
\caption{Ternary bit write pattern generating encoder placed before Multi-Level Memory for writing analog input voltage}
\label{SystemOverview}
\end{figure*}

Human memory system requires encoding the information that comes from sensory inputs in the form it can process and store. The information can be encoded into visual, acoustic and semantic representations\cite{mcleod2007stages}. Information transmission and transformation of such character is achieved with the neural network of the brain \cite{berlucchi2002origin}.

First, sensory neurons receive the signals in the analog domain and pass the signal to the neurons of the network that maintains a voltage gradient across its membrane, that has a different charge, depending on the ions within the cell \cite{llinas1988intrinsic}. If the voltage changes significantly, an electrochemical pulse called an action potential (or nerve impulse) is generated. Thus the network of such neurons interacts between themselves in the mixed signal domains. \cite{rojas2013neural}. The encoded information is then stored in form of the synapses - a neuron to neuron connection, which strength can be increased or decreased over time and learning process. Every single neuron can form thousands of connections with other neurons in this way, which means a typical brain would have 100 trillion synapses \cite{fauci2008harrison}. This information according to neuroscience literature is stored in form of analog information in the human brain. 

Data processing that takes place in human brain outperforms modern processors on many tasks like data classification and pattern recognition. This inspires neuromorphic engineers to build massively
parallel architectures emulating the human brain by modeling low-power computing elements - neurons and adaptive memory elements -synapses.  One of the ways to implement computing systems that can mimic massive parallelism and low power operation as in brain is to scale dense non-volatile memory crossbar arrays \cite{eryilmaz2014brain}. In this paper, a neuron inspired memristor based multi-level memory (MLM) with analog data encoder that can be used in crossbar arrays is presented. We build the motivation of using multi-level memories on the premise that analog memories are integral to the development of cognitive algorithms, both at sensory processing level and in high levels of cognitive function implementations. The presented circuit is configured to store 10 discrete analog values and simulated for different temperature conditions.

The paper is organized as follows: Section II provides background information of the memory cell used in the paper. Further, the circuit for data encoder that generates control voltages for writing discrete analog values is discussed. In following section simulation results of the memory cell with data encoder at different temperature conditions is presented.

\section {Background}
In this paper, we present an encoder for proposed memory cell in\cite{irmanova2017multi} based on memristors, arranged in a potential divider configuration that could emulate a weighted addition that is indicative of dendritic segments of the neuron. To enable writing discrete analog values into the cell it is required to encode the analog input into ternary bit write pattern control voltages that are used to program memristors in the sub-cells. The memory cell is programmed using ternary bits of different amplitude [$0V;2.5V;4V$] of corresponding logical values$[0;1;2$]. Fig. 1 Shows the example of writing analog input voltage of an amplitude 1.3V which correspond to $012$ write pattern that generates [$0V;2.5V;4V$] write pulses to be sent to the relative [$V_{w1}; V_{w1}; V_{w1}$] write ports.

\begin{figure}[!ht]
\centering
\includegraphics[width=6cm]{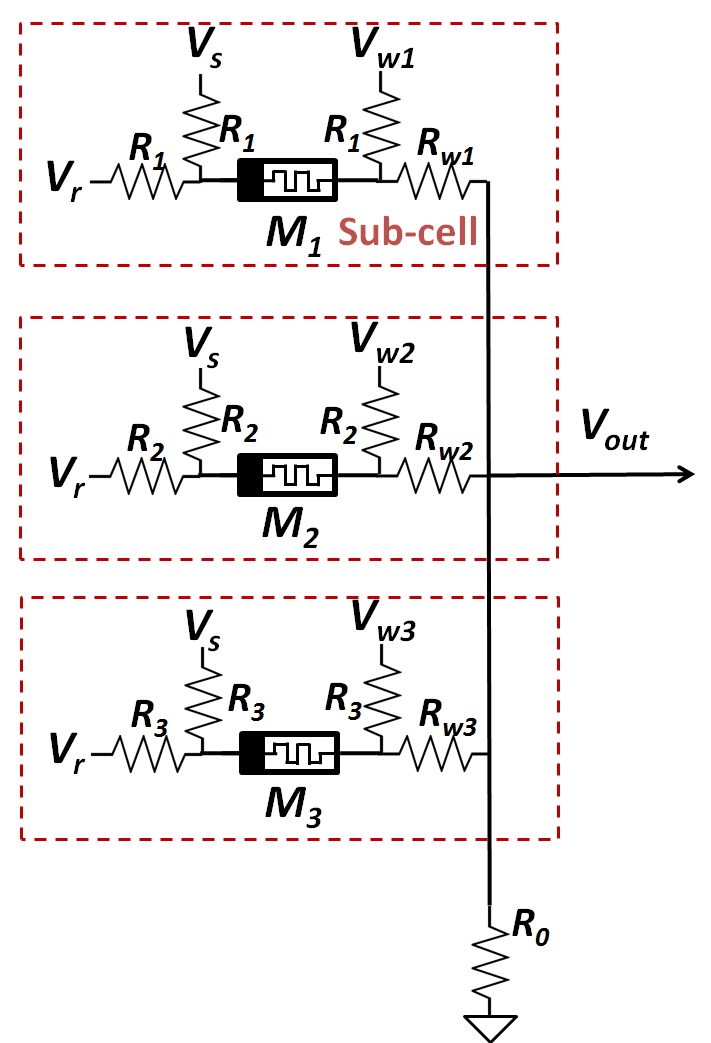}
\caption{{Circuit design of the multi-level memory cell}}
\label{read}
\end{figure}

\begin{table}[]
\centering
\caption{Write pattern for corresponding input range}
\label{my-label}
\begin{tabular}{|l|l|l|l|l|}
\hline
\# & Input range                                                       & $V_{w1}$ & $V_{w3}$ & $V_{w3}$ \\ \hline
1  & \begin{tabular}[c]{@{}l@{}}$a_1$ = 0\\ $a_2$= 0.3\end{tabular}    & 2        & 2        & 2        \\ \hline
2  & \begin{tabular}[c]{@{}l@{}}$a_1$ = 0.31\\ $a_2$= 0.6\end{tabular} & 1        & 2        & 2        \\ \hline
3  & \begin{tabular}[c]{@{}l@{}}$a_1$ = 0.61\\ $a_2$= 0.9\end{tabular} & 1        & 1        & 2        \\ \hline
4  & \begin{tabular}[c]{@{}l@{}}$a_1$ = 0.91\\ $a_2$= 1.2\end{tabular} & 0        & 2        & 2        \\ \hline
5  & \begin{tabular}[c]{@{}l@{}}$a_1$ = 1.21\\ $a_2$= 1.5\end{tabular} & 0        & 1        & 2        \\ \hline
6  & \begin{tabular}[c]{@{}l@{}}$a_1$ = 1.51\\ $a_2$= 1.8\end{tabular} & 1        & 1        & 1        \\ \hline
7  & \begin{tabular}[c]{@{}l@{}}$a_1$ = 1.81\\ $a_2$= 2.1\end{tabular} & 0        & 0        & 2        \\ \hline
8  & \begin{tabular}[c]{@{}l@{}}$a_1$ = 2.11\\ $a_2$= 2.4\end{tabular} & 0        & 1        & 1        \\ \hline
9  & \begin{tabular}[c]{@{}l@{}}$a_1$ = 2.41\\ $a_2$= 2.7\end{tabular} & 0        & 0        & 1        \\ \hline
10 & \begin{tabular}[c]{@{}l@{}}$a_1$ = 2.71\\ $a_2$= 3\end{tabular}   & 0        & 0        & 0        \\ \hline
\end{tabular}
\end{table}


To start with, the main advantage of the design of given memory cell is that it is purely based on memristors. As memristors are nanoscale devices that operate with current leakage which will result in less area and power consumption \cite{sarwar2013memristor}. 
In proposed memory cell shown in Fig. 2, the ungrounded node serves as an input port for receiving signals and represents the membrane resistance.  Its structure of voltage divider provides with different $V_{out}$ levels across $R_0$ during the read operation, resulting from the $V_r$ read input voltage applied across the membrane resistance, where the membrane resistance is represented as the total resistance of memristors connected in parallel. The cell is programmed to ternary logic to achieve an increased number of states by exploiting the state changes of multiple memristors at a given time, overcoming the limitations of using single memristor that is practically limited by the device variability and implementation complexity. Each memristor of the cell is placed into sub-cells, which are used for programming the memristor to high $V_2$, medium $V_1$ and low $V_0$ states. To program each memristor to the desired state the write signal $V_w$ is applied through the write port from the positive terminal of the memristors, and the reset signal $V_s$ that precedes every write operation to erase previous states is applied from the negative terminal - reset port of the device. In general, the memory cell with $n$ sub-cells can store $mn$  discrete values, where $m$ is the number of different levels of voltage that is applied through $Vw_1, Vw_2, Vw_3$ write ports of the sub-cells. In this paper, simulation results of the memory cell with $n=3$ sub-cells are presented, each of them programmable to $m=3 (V_0, V_1, V_2)$ states which could result in up to 27 level discrete analog memory. It is to be noted that resistance values used for connecting to reset, write, read ports $R_1, R_2, R_3$ should be set as R 1≠ R2 ≠ R3, otherwise, for $n=3$ and $m=3$ memory cell, discrete output levels will decrease down to 10 different values. This is the result of equal voltage drop within each sub-cell shown in Fig.2 which combines to the overall number of output states.  Achieved 10 $V_{out}$ output states of the memory cell are presented in Fig. 5. For reading the $V_{out}$  output states of the circuit read signal $V_r$ is applied to the $V_r$ read ports of the sub-cell. For the setup with an encoder, the timeline of the single cycle of resetting, writing, and reading is provided in Fig. 4. For simulations of the circuit initially $4V$ write signal $V_1$ corresponding to logic 2 , $2.5V$ write signal $V_1$ corresponding to logic 1 and $0V$ write signal $V_0$ corresponding to logic 0; $4V$ $Vs$ reset signal; $0.05V$ $Vr$ read signals were used. As regards resistors values,in this paper we used the setup of  $R_1 = R_2 = R_3$  resistances was set as $500 \Omega$, and $R_{w1} = R_{w2} = R_{w3}$ resistances was set $1500 \Omega$ , respectively. The resistance of $R_w$ write port is greater as it connects to the encoder output ports and to ensure the working of the memory is stable the load resistance was increased. For the simulations, the memory cell performance memristor model described in \cite{pickett2009switching} was used in LTSpice \cite {ltspice2013linear}.

\begin{figure}[!ht]
\centering
\includegraphics[width=10cm]{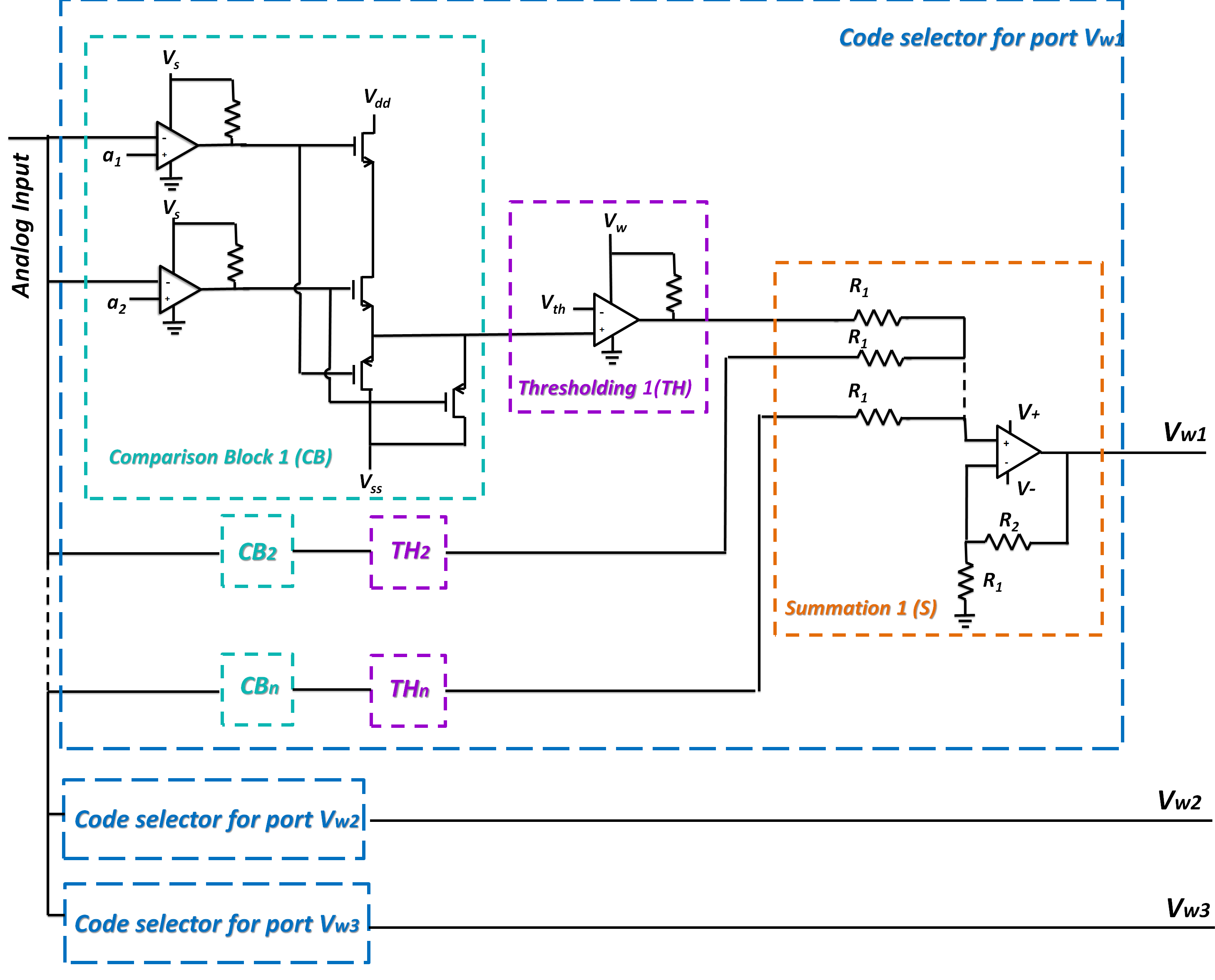}
\caption{{Circuit design of an encoder for MLM}}
\label{read}
\end{figure}

\begin{figure}[!ht]
    \centering

      \subfigure[]{
    \includegraphics[width=6cm]{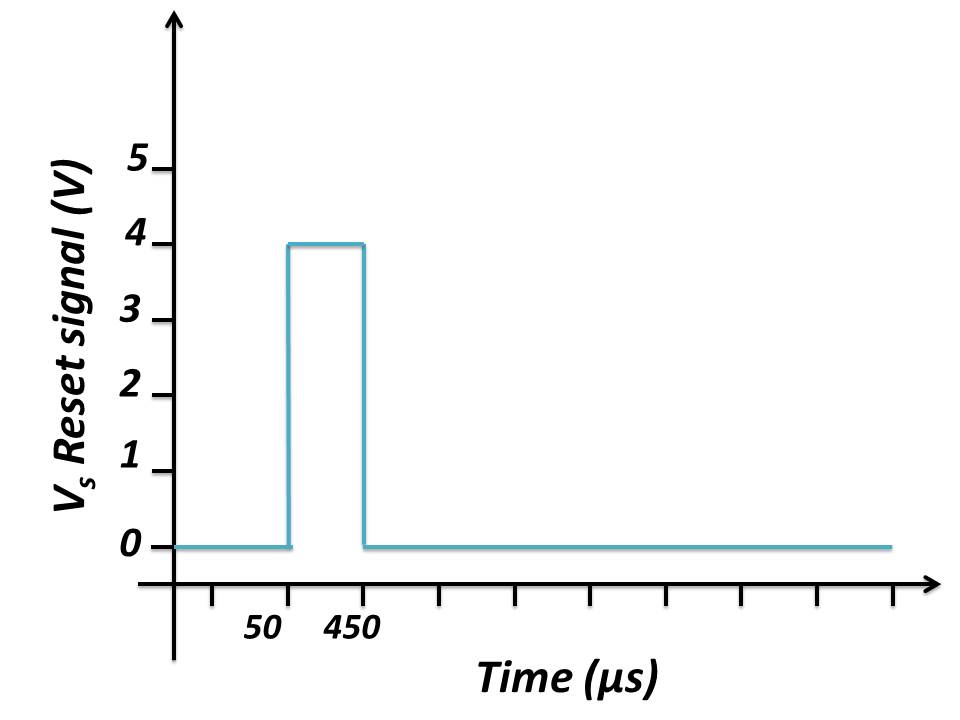}}
      \subfigure[]{
    \includegraphics[width=6cm]{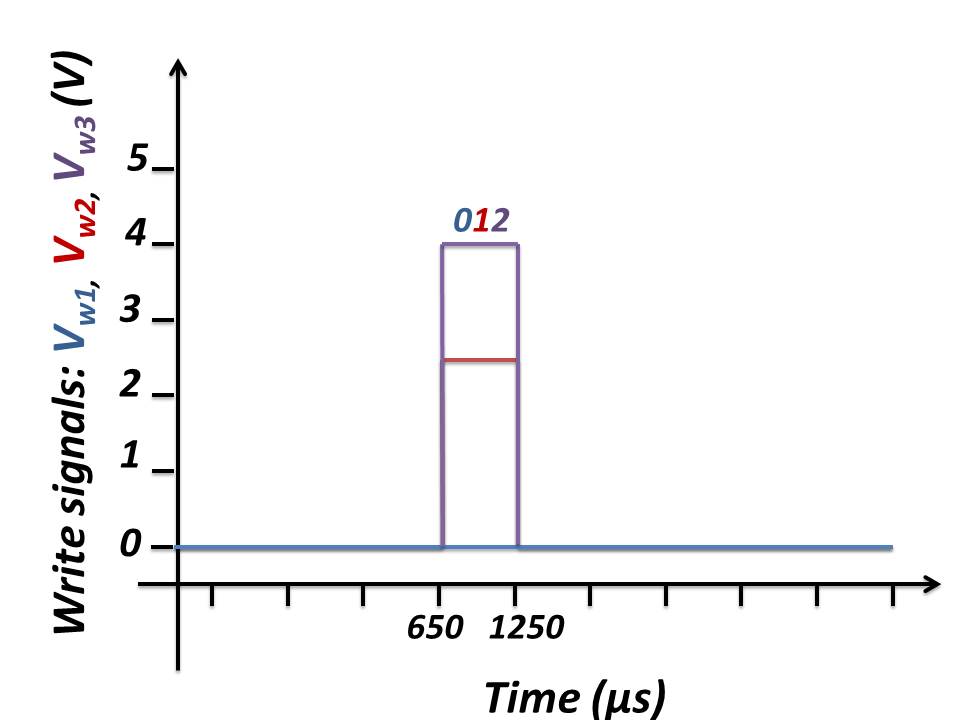}}
    \subfigure[]{
    \includegraphics[width=6cm]{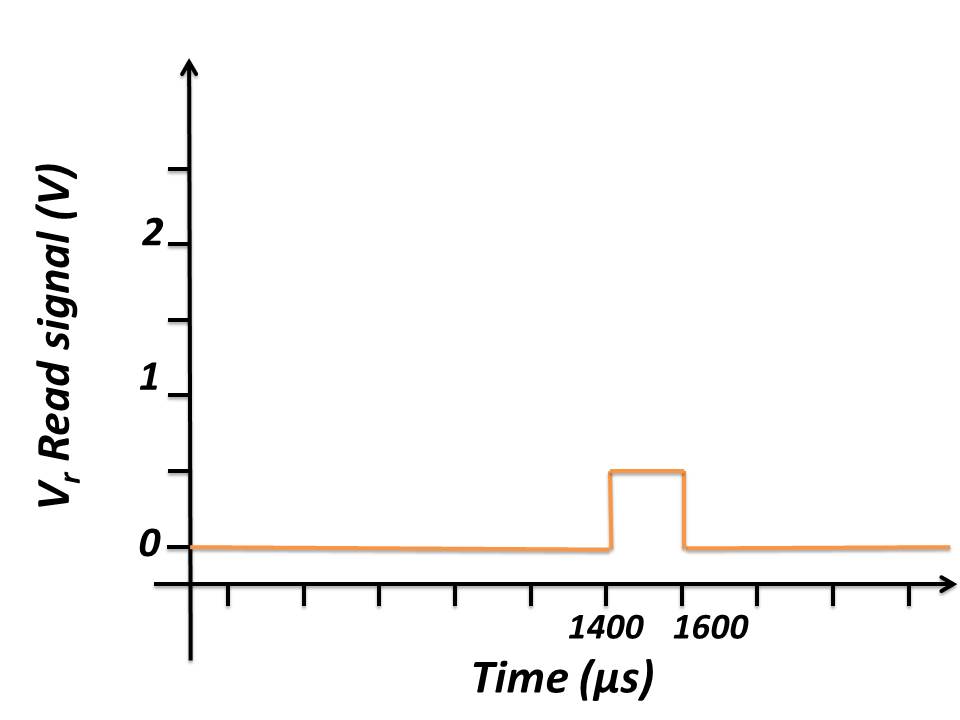}}

    \caption{Time diagram of a single cycle of the (a) reset (b) write and (c) read signal }
       \label{td11}
\end{figure}

\begin{table}[]
\centering
\caption{Encoder circuit configuration}
\label{my-label}
\begin{tabular}{|l|l|}
\hline
\begin{tabular}[c]{@{}l@{}}NMOS:\\  W/L ($\mu$m) \end{tabular} & 0.36/0.18                                          \\ \hline
\begin{tabular}[c]{@{}l@{}}PMOS:\\  W/L ($\mu$m) \end{tabular} & 0.72/0.18                                          \\ \hline
$V+/V-$            & 3V/-3V                                             \\ \hline
$V_dd$             & 1.5V                                               \\ \hline
$V_{ss}$           & -1.5V                                              \\ \hline
$V_{th}$           & 0.3                                                \\ \hline
$V_{s}$            & 4V                                                 \\ \hline
$V_{w}$            & 2.5V, 4V \\ \hline
$R_1$              & 10k                                                \\ \hline
$R_2$              & 10(n-1)k                                           \\ \hline
\end{tabular}
\end{table}

\begin{figure*}[!ht]
    \centering
      \subfigure[]{
    \includegraphics[height=5cm]{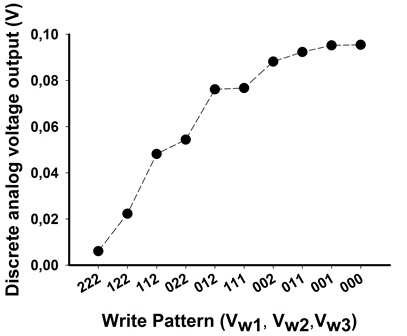}}
    \subfigure[]{
    \includegraphics[height=5cm]{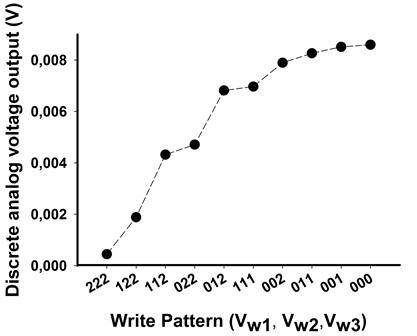}}
 
    \caption{{Achieved discrete analog levels of 10 level memory cell (a) directly applying control voltages of write pattern and (b) using an encoder for generating the write pattern}}
\end{figure*}

\begin{figure*}[!ht]

\centering
	\subfigure[]
    {
	\includegraphics[ height=5cm ]{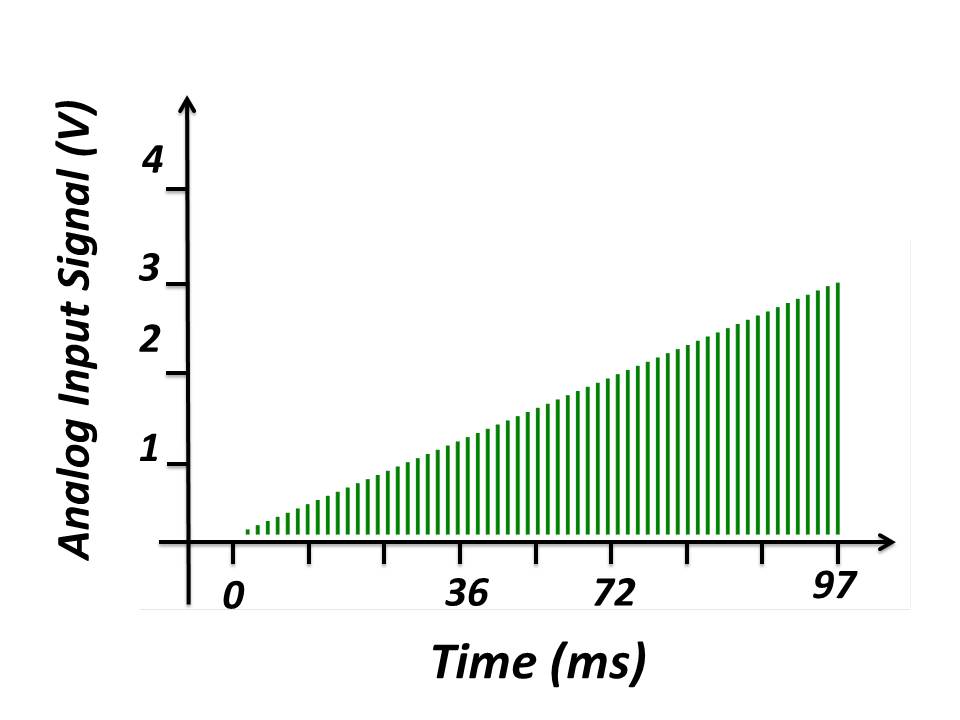}
	\label{l10}
    }
    \subfigure[]
    {
    \includegraphics[ height=5cm]{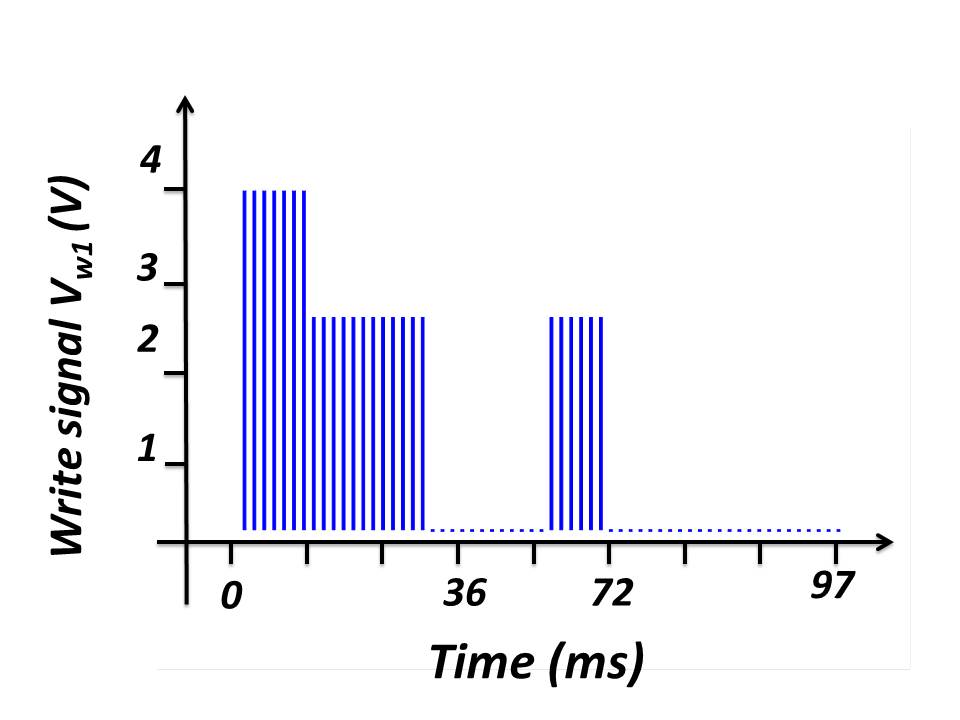}
  	 \label{l27}
     }
     \subfigure[]{
    \includegraphics[ height=5cm]{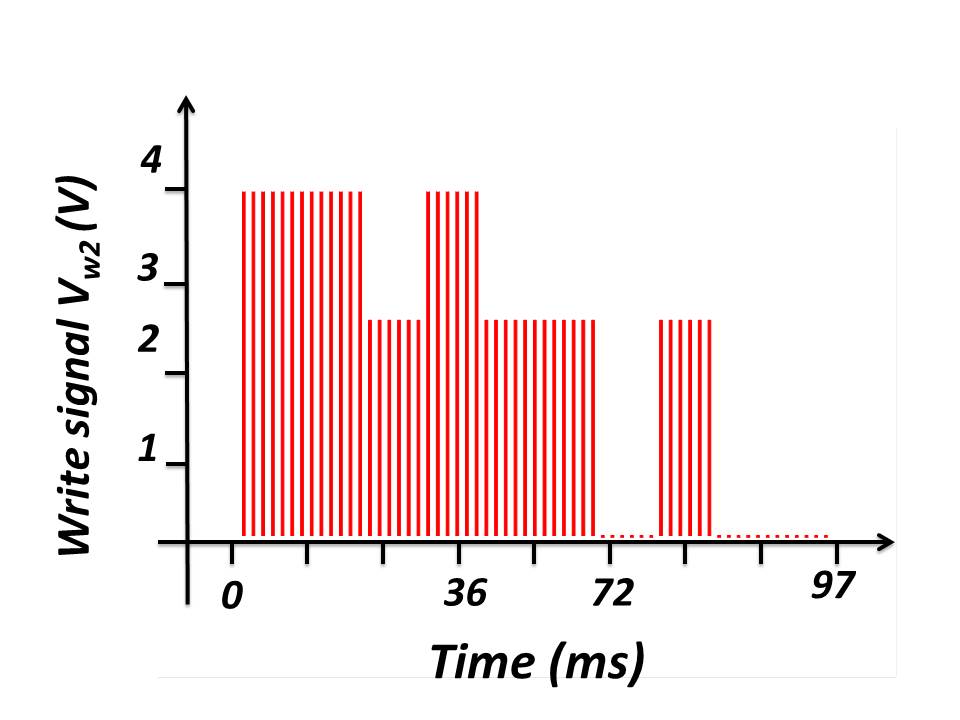}
  	\label{l243}
5    }
     \subfigure[]
     {      
   \includegraphics[height=5cm]{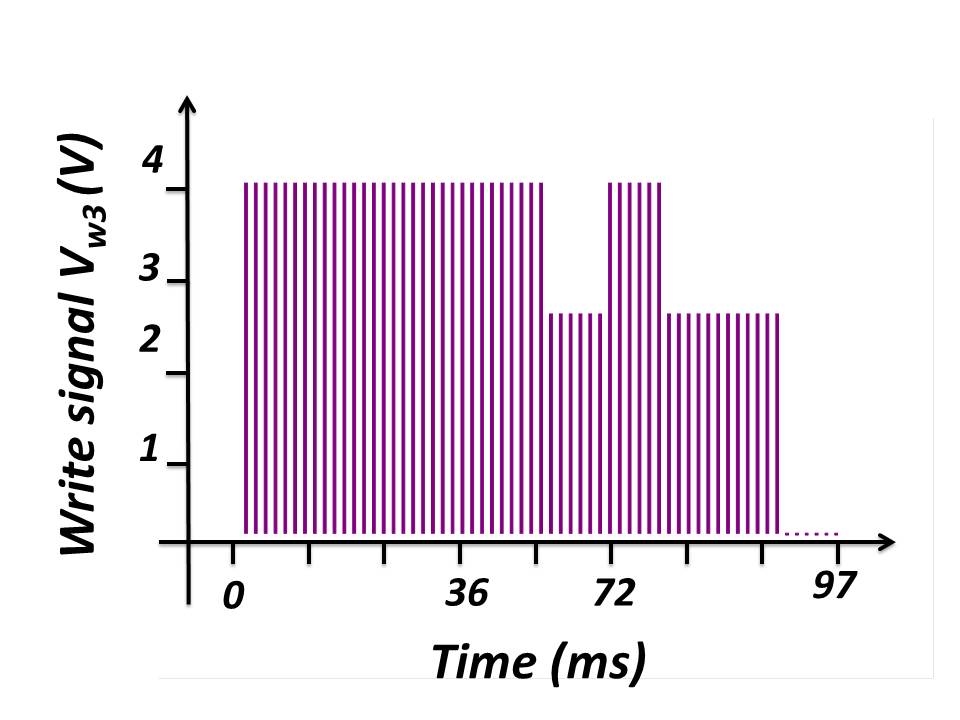}
   \label{l1024}
   }
     \caption{Generated output write pattern signal from (a) analog input signal fed to the encoder ranged from $0V$ to $3V$ with $0.05V$ step for (b) $V_{w1}$, (c) $V_{w2}$,  (d) $V_{w3}$ write ports.}
     \end{figure*}
\begin{table*}[]
\centering
\caption{Discrete analog output voltages of 10 level memory cell with data encoder at different temperature condition}
\label{my-label}
\begin{tabular}{|c|c|c|c|c|c|c|c|c|}
\hline
\multirow{2}{*}{} & \multicolumn{2}{c|}{$V_{out}$ of MLM at $20^{\circ}C$} & \multicolumn{2}{c|}{$V_{out}$ of MLM at $30^{\circ}C$} & \multicolumn{2}{c|}{$V_{out}$ of MLM at $40^{\circ}C$} & \multicolumn{2}{c|}{$V_{out}$ of MLM at $50^{\circ}C$} \\ \cline{2-9} 
                  & Mean $(V)$                       & STDEV                     & Mean $(V)$                       & STDEV                     & Mean $(V)$                        & STDEV                     & Mean $(V)$                       & STDEV                     \\ \hline
222               & 4.462E-04                  & 1.079E-06                 & 4.461E-04                  & 1.130E-06                 & 4.461E-04                  & 9.944E-07                 & 4.461E-04                  & 1.060E-06                 \\ \hline
122               & 1.884E-03                  & 5.746E-07                 & 1.884E-03                  & 6.277E-07                 & 1.884E-03                  & 6.277E-07                 & 1.883E-03                  & 4.765E-07                 \\ \hline
112               & 4.319E-03                  & 4.664E-07                 & 4.319E-03                  & 5.979E-07                 & 4.319E-03                  & 5.979E-07                 & 4.319E-03                  & 6.073E-07                 \\ \hline
022               & 4.712E-03                  & 2.295E-05                 & 4.712E-03                  & 2.326E-05                 & 4.712E-03                  & 2.326E-05                 & 4.712E-03                  & 2.320E-05                 \\ \hline
012               & 6.816E-03                  & 8.382E-06                 & 6.816E-03                  & 8.310E-06                 & 6.816E-03                  & 8.326E-06                 & 6.816E-03                  & 4.700E-07                 \\ \hline
111               & 6.967E-03                  & 6.282E-06                 & 6.969E-03                  & 8.235E-06                 & 6.970E-03                  & 8.321E-06                 & 6.970E-03                  & 7.565E-06                 \\ \hline
002               & 7.898E-03                  & 4.084E-05                 & 7.896E-03                  & 4.041E-05                 & 7.896E-03                  & 4.013E-05                 & 7.896E-03                  & 3.864E-03                 \\ \hline
011               & 8.266E-03                  & 1.351E-05                 & 8.266E-03                  & 1.342E-05                 & 8.266E-03                  & 1.342E-05                 & 8.266E-03                  & 1.342E-05                 \\ \hline
001               & 8.513E-03                  & 2.829E-05                 & 8.513E-03                  & 2.798E-05                 & 8.513E-03                  & 2.798E-05                 & 8.513E-03                  & 8.513E-03                 \\ \hline
000               & 8.598E-03                  & 2.015E-05                 & 8.598E-03                  & 2.017E-05                 & 8.598E-03                  & 2.017E-05                 & 8.598E-03                  & 1.969E-05                 \\ \hline
\end{tabular}
\end{table*}

\section{Encoder design}
To design an encoder for the proposed memory cell, first, its 10 discrete output levels were sorted from small to large and its corresponding write pattern codes (Fig. 5) were assigned to the analog value ranges [$a_1; a_2$] that are given in Table 1. In this paper, we assume that analog input signal variation will be within 3V amplitude and this value is divided to 10 ranges.
To encode analog input signal into write pattern $V_{w1}, V_{w2}, V_{w3}$ control voltages it is required to identify which  range [$a_1; a_2$] of analog values it belongs to and after that generate the voltage pulses according to its assigned write pattern code [$0V; 2.5V; 4V$] for corresponding [$0; 1; 2$] logic values. To implement this it was decided to construct separate \textit{code selector blocks} for each write ports $V_{w}$. The circuit design of this encoder is presented in Fig. 3. The circuit consists of 3 code selecting blocks for 3 write ports $V_{w1}, V_{w2}, V_{w3}$ of the memory cell. Selector itself consists of comparison, thresholding, and summation blocks. At the comparison block outputs of 2 comparators with reference voltages [$a_1; a_2$] are fed to the CMOS AND gate. In case the output of this block is high it will generate assigned $V_w$ voltage pulse from the thresholding block which consists of a comparator with $V_th$ reference voltage. As the input signal can belong to only one range  [$a_1; a_2$], outputs of all thresholding blocks are summed to implement logical conjunction and produce $V_{w}$ write signal. For comparison and thresholding blocks the model of operational amplifier $LTC6702$ and for summing amplifier at the summation block model $LTC1006$ was used. While AND gate circuit was built with $0.18\mu$ CMOS transistor technology. Further information on circuit configuration is provided in Table 2.

\section{Results}

To test the performance of the encoder train of pulses with an amplitude ranged from $0V$ to $3V$ with the step of $0.05V$ and the duration of 0.6ms was fed to the circuit input port (Fig.6 (a)). Before writing the signals from given input train, the memory cell was reset with $4V$ amplitude signal, preceding each write operation as shown in Fig.4. Afterwards, the write signal is sent to the encoder and produced write pattern control voltages was fed to $V_{w1}, V_{w2}, V_{w3}$ write ports for programming the memory cell to discrete analog level. Then the read signal with the duration of 0.2ms is sent to check the $V_{out}$ output voltage. Output of the encoder resulted in [$4V;2, V;(-0.2;0.004)V$] for [$2;1;0$] logical values relatively. Slight variations in the write pattern resulted as well in the shift of $V_{out}$ output levels of the memory cell. This can be seen in Fig. 5(a)  and Fig. 5(b). In Fig. 5(a) the outputs of the memory cell were programmed by directly applying write pattern voltages, without an encoder, while for Fig. 5(b) the output of the memory cell operating with an encoder at the temperature of $50^{\circ}C$  is provided. Table 3 shows average output levels of the memory and its standard deviation at the temperature of  $20^{\circ}C$,  $30^{\circ}C$,  $40^{\circ}C$ and  $50^{\circ}C$. It can be seen that the output of the memory cell programmed with an encoder change negligible at this range of temperature. Power dissipation for single data encoder an memory cell pair can reach up to 94.5mW.

\section{Conclusion}
In this paper, we presented the multi-level memristive memory cell with data encoding control circuits. Presented memory cell can be used in the crossbar arrays to store discrete analog values. The proposed encoder circuit consists of operational amplifiers and CMOS transistors which configuration can be further improved to reduce power and area consumption. Simulation results showed that the memory cell outputs negligibly vary at different temperature conditions. Inspired from neuron structure the memory cell stores discrete analog values encoding the analog input, which can be used in the implementations of the neuromorphic systems for synaptic weights storage. Incorporating analog storage of discrete values allows analog data computation which results in improved performance in terms of speed. To ensure the compatibility of the memory cell with data encoder and stable performance of the memory cell resistance values of the resistive network of sub-cells were increased which involved increasing write and reset control voltage amplitudes. 

\bibliographystyle{unsrt}
\bibliography{ref.bib}



\end{document}